# Educational cosmic ray experiments with Geiger counters

F. BLANCO, F. FICHERA, P. LA ROCCA, F. LIBRIZZI, O. PARASOLE AND F. RIGGI

*INFN and Department of Physics and Astronomy, University of Catania, Via S.Sofia 64, I-95123 Catania, Italy*

**Summary.** — Experiments concerning the physics of cosmic rays offer to high-school teachers and students a relatively easy approach to the field of research in high energy physics. The detection of cosmic rays does not necessarily require the use of sophisticated equipment, and various properties of the cosmic radiation can be observed and analysed even by the use of a single Geiger counter. Nevertheless, the variety of such kind of experiments and the results obtained are limited because of the inclusive nature of these measurements. A significant improvement may be obtained when two or more Geiger counters are operated in coincidence. In this paper we discuss the potential of performing educational cosmic ray experiments with Geiger counters. In order to show also the educational value of coincidence techniques, preliminary results of cosmic ray experiments carried out by the use of a simple coincidence circuit are briefly discussed.

## 1. — Introduction

In recent years examples of collaborations which involve high schools, university and research centres in projects concerning cosmic ray physics have become more and more numerous all over the world [1,2]. The main reason of this trend lies in the evident educational aspects that characterize experiments for the detection of cosmic rays: this kind of activities offers to school-teachers and associated groups of students the possibility to be trained in the world of experimental high energy physics, since not only they may learn about the basics of cosmic ray physics, but also stay in touch with several aspects of a research activity.

Detection configurations usually employed for these projects consist of scintillators, mostly arranged as telescopes or small arrays. However, if students may derive some benefit from the use of sophisticated equipments, it is possible to examine several properties of the cosmic radiation even by the use of simple Geiger counters. This solution inevitably implies a reduction in the counting rate due to the small size of such devices. On the other side, through the use of Geiger counters and associated simple electronics and data acquisition, schools have the possibility to carry out cosmic ray experiments in a completely autonomous way, because the use of such apparatus does not require sophisticated skills on detection techniques and also runs even within the limited budget of a high school.

In this paper, we briefly describe a set of educational cosmic ray experiments which are part of an established collaboration between the INFN, the Department of Physics and Astronomy of our University and local area high schools. Both inclusive and coincidence data have been obtained in the framework of these experiments. Section 2 reports a summary of inclusive experiments carried out with single Geiger counters, which are within a high-school's reach. The potential of such experiments may be largely improved by the use of coincidence techniques. Some example of coincidence experiments with small Geiger counters is reported in Section 3. Finally, Section 4 draws some conclusions and outlook.

## 2. — Examples of inclusive measurements performed with single Geiger counters

In most of the experiments discussed in this paper, commercial Geiger counters and associated electronics/data acquisition have been used. A typical arrangement is sketched in fig. 1.

This equipment is composed of a PASCO Interface (for instance Mod. 500 or 750) and various digital or analog sensors which may be connected to it. Data are collected and stored by means of a user-friendly software, so that the whole represents a useful ready-made system for experiments in physics [3]. Similar systems, based on the use of on-line sensors, are available on the market.

Despite its evident simplicity, the experimental set-up of fig. 1 allows to perform various measurements of educational interest. For instance, one of the simplest investigations which may be performed is the study of the time variations in the cosmic ray flux at the sea level. Such fluctuations may be caused by several factors and are related to phenomena which involve the Sun, the Earth and its environment in space.

A variation that is often studied is the "barometric effect": owing to the mass absorption provided by the Earth's atmosphere, variations of the atmospheric pressure result in small fluctuations of the cosmic ray flux. The barometric effect is estimated by the use of the barometric coefficient $\beta$, which expresses the percentage variation in the cosmic ray intensity caused by a pressure change of 1 mbar. The value of $\beta$ depends however on the particular experimental set-up used for the detection, on its geographical location and its local environment. The barometric coefficient is also strongly dependent on the nature of the secondary component being detected: for instance the value of $\beta$ for the neutron component is about 0.7%/mbar, while it is smaller for the ionizing component ($\beta\sim$0.1-0.2%/mbar). Especially for the ionizing component, we are dealing with very small variations and a careful study of such effect should require a high statistical accuracy. The use of detectors with small area (such as the Geiger counters previously proposed) is not recommended for such kind of analysis. However, a pilot measurement carried out by the use of a single Geiger counter showed that it is possible to obtain a realistic value of the barometric coefficient even by means of detectors with a low counting rate [4, 5]. For such investigation we used a Geiger counter together with a barometric sensor for a parallel recording of the local pressure. Owing to the low Geiger counting rate (about 0.3 Hz), the atmospheric pressure and the counting rate were monitored continuously for a period of approximately 100 days in 30' steps. Data analysis of such measurements offers a good opportunity to use statistical methods. For example, periodic variations of the local atmospheric pressure may be studied by extracting a correlogram from the data [4]. This gives a quantitative measure of the correlation between the pressure at the time t and that at the time t+Δt, as a function of the time interval Δt. Figure 2 shows the correlogram from the pressure time series: the well-known periodicity of 24 hours emerges clearly from the plot.

The value of the barometric coefficient β may be extracted through a standard weighted fit of the measured (count rate, pressure) pairs. For small pressure changes we may consider the relation between the counting rate R and the atmospheric pressure P to be linear and in this approximation our data gave a barometric coefficient of β=(0.023±0.009)%/mbar. However, more in-depth analyses may be performed. For instance, in order to minimize the effect of daily variations not connected to pressure changes, the linear fit may be performed considering the daily averages of pressure and counting rate. Besides daily fluctuations, the cosmic flux may be also affected by the presence of long term variations, which can destroy the correlation between R and P. An accurate analysis of the overall data led to the conclusion that in the second half of the measurement period, cosmic flux suffered an increase not due to the atmospheric pressure. This variation is visible, only in a qualitative manner, in fig. 3, that shows the correlation plot between daily averages of pressure and counting rate for the overall set of data (upper) and for the first and the second half of the measurement period (middle and lower respectively). To take into account such long-term variations, we studied the correlation within small time periods (10 days), obtaining an average barometric coefficient β=(0.051±0.015)%/mbar.

The search for long term variations may be facilitated by comparing the measured data with those taken by professional stations for the detection of cosmic rays [6]. For example, data from the Moscow Neutron Monitor confirm the presence of a variation (about 1 %) of the cosmic flux in the second part of the period under consideration. So another didactic exercise could be the study of the correlation between data measured with a Geiger counter and fluxes of neutron and muons obtained by professional monitor stations during a Forbush decrease. In fact, besides the atmospheric effects, a lot of changes in cosmic ray intensity are due to the solar activity. The Forbush decrease belongs to this class: it is a rapid, short-term variation resulting from solar flares and mass ejections of the Sun corona. This effect is more difficult to see with muon detectors since it affects mainly the low energy particles. Data measured with a Geiger counter during a Forbush decrease [6] were compared to fluxes reported by two professional neutron stations: in both cases, the statistical errors did not allow to directly see a correlation between the muon flux in the Geiger counter and the neutron flux. However we must remind that Forbush events are characterized by two different phases, a first rapid intensity reduction, followed by a slow recovery. Therefore, it is possible to study the degree of correlation during these different phases, using both the data measured with the Geiger counter and data measured with a larger statistics from a professional muon telescope.

The solar activity affects the flux of cosmic rays not only through exceptional events (such as the Forbush decrease) but also through periodic effects (such as the 11-years, annual, 27-days, solar day cycles). For instance the data collected for the study of the barometric effect were again analyzed to search for daily variations [7]. The amplitude and the phase of such variations are not constant over long periods and depend on the location of the detecting apparatus, as well as on the nature of the cosmic component being detected. Once the data were corrected for the barometric effect, we normalized the hourly flux to the average daily flux so as to minimize the effect of long-term variations. Carrying out an harmonic analysis for each individual day, we found a mean amplitude of 0.16% and a mean phase $\phi=22.3$ hours (solar local time). An alternative analysis was performed by considering the ratio between the mean hourly flux in a selected interval along the day and the corresponding mean flux in the rest of day. In fig. 4 we choose a time interval of 12 consecutive hours and this ratio is plotted by shifting the centroid of the selected time interval in steps of 30 minutes. Both analyses, carried out for a relatively long period (about 100 days), led to evidences of a periodic behaviour of the flux along the day.

The interpretation of all the effects previously described is not however straightforward, particularly because of the inclusive nature of the observations. These investigations must then be considered as pilot studies in order to introduce high-school teams to several aspects of a physic research activity, from detection techniques to problems in data analysis and interpretation.

### 3. — Use of coincidence techniques in educational cosmic ray experiments

In cosmic ray physics the use of coincidence techniques is a very common practise. Inclusive measurements often lead to results difficult to interpret; a partial improvement may be obtained by means of coincidence techniques, because they allow to fix more restrictive work conditions so as to select only particular classes of events: for example, a telescope of two detectors in coincidence defines a particular orientation, allowing to better control the effects of irregularity due to materials surrounding the telescope. Moreover, the use of coincidence techniques gives the possibility to investigate several aspects of the cosmic radiation which cannot be explored simply by means of inclusive measurements.

For most of these investigations, people have generally used relatively large area plastic scintillators, in order to have an acceptable rate of coincidences. However, fast detectors require complex and expensive electronics, which is not compatible with the limited resources of a high-school. On the contrary, inexpensive coincidence circuits for Geiger counters may be easily built-up even by a high-school student, since signals produced by these detectors allow to use simple AND gates.

Coincidence circuits for Geiger counters are not easily available on the market and even companies which provide systems of on-line sensors do not make available such kind of devices either. In order to use the same system in fig. 1, we have devised a simple coincidence circuit which provides a signal compatible with the PASCO interface. The basics of this circuit is elementary: the signals of two Geiger counters are firstly sent to a dual retriggerable monostable multivibrator, in order to provide standard signals, and then are used as inputs for an AND gate. Actually the monostable multivibrator is not necessary if we use the Pasco Geiger sensors, which provide a TTL-like signal of amplitude + 5 V and width around 120 µs. This solution, however, gives the possibility to use the circuit also with other kinds of Geiger sensors. A detailed description of the electronics used is reported in Ref. [8].

Once tested, the system has been used to carry out some educational coincidence experiments. A typical investigation which requires the use of a coincidence circuit is the study of the angular dependence of the cosmic flux. For this measurement we arranged two Geiger counters in order to set up a telescope for cosmic rays. The coincidence between the two detectors allows to select only those particles which simultaneously pass through both of them, so as to roughly select a particular direction for the incoming radiation. The solid angle of the telescope and, hence, its counting rate, are determined by the size of each detector and by their distance: the more the Geiger counters are distant, the more the telescope orientation is precise and its counting rate low. This arrangement allows to study the dependence of cosmic ray flux at the sea level on the zenithal ($\theta$) or azimuthal ($\varphi$) angle. Owing to the absorption effect of the Earth's atmosphere, the expected cosmic flux varies with a $\cos^2\theta$ trend for the zenithal angle, while it is quite uniform for the azimuthal angle. Figure 5 shows the number of coincidences per minute as a function of $\cos^2\theta$, when the two Geiger counters are placed at a distance of 10 cm. Within the experimental errors, the points should lie on a straight line if the distribution follows a $\cos^2\theta$ law, as it appears from the weighted fit to the data (solid line). Due to the finite size of the counters, and their mutual distance, the effective angle defined by the telescope geometry is different from the nominal one. In order to evaluate the effective angle $\theta_{eff}$ for each detection geometry, a Monte Carlo simulation of the possible trajectories crossing the two Geiger cylindrical volumes (radius 1 cm, length 9 cm) was performed. Figure 6 (left) shows the distribution of the polar angles for a nominal orientation $\theta = 40°$ and a mutual distance of 10 cm. The centroid of the distribution ($10^5$ events) was taken as an estimate of the effective angle, with an uncertainty given by the RMS of the distribution. Figure 6 (right) shows the correlation between the nominal and the effective angle, for the geometry used in the present set-up. As it can be seen, detection angles close to $\theta = 0°$ (vertical orientations) are difficult to obtain, due to the sizes of the counters, which allow also for inclined tracks even at $\theta = 0°$, unless very large relative distances between the two counters are selected. On the other side, at large angles, the effective angle is very close to the nominal one.

A similar set of measurements was carried out to investigate the azimuthal angle distribution. Once fixed the zenithal angle $\theta$, we expect (and actually observe) a nearly uniform distribution for the angle $\varphi$. Small variations may be observed only in the presence of huge materials around the telescope (such as a close mountain), but this effect is very difficult to observe with such apparatus, since a very high statistics is required. Precision experiments should also allow to investigate the small east-west asymmetry, due to the charge of the particles being detected.

Another typical coincidence experiment is the measure of the decoherence curve. For such investigation two or more detectors are placed on the same horizontal plane, separated by some distance; the decoherence curve is the dependence of the coincidence rate upon the detector separation. The detected coincidences are due in such case to different cosmic particles, which however belong to the same air shower. Professional measurements of such curve show that the coincidence rate decreases dramatically as the separation of detectors increases. For such reason, the use of Geiger counters requires long running times to get adequate statistics. Moreover, the two Geiger counters cannot be separated too much, because at large distances the counting rate becomes comparable with that of the spurious coincidences $N_s = 2 \tau N_1 N_2$, where $N_1$, $N_2$ are the individual count rates of the two detectors and $\tau$ is the coincidence window. For typical individual rates $N_1 = N_2 = 0.3$ Hz and $\tau = 150$ µs, the spurious rate is in the order of 1 coincidence in 10 hours. Figure 7 shows a preliminary result on the decoherence curve obtained with two small Geiger counters. Data were corrected for spurious coincidences, since their influence is not negligible. The observed trend follows approximately a $r^{-n}$ shape, with $n \cong 2$. We must say that these results are very preliminary, since a higher statistics and a careful analysis of the influence of any shield above the counters are required. Moreover, this arrangement allows to also have coincidences due to single nearly horizontal particles, which pass through both detectors. Even if very rare, such events could alter the effective coincidence rate, particularly at large distances. A possible solution could be putting a third detector above one of the two counters and selecting triple coincidences: in this way we could be confident that the coincidences are given by two different particles, since a single particle cannot pass through all three detectors.

A lot of other educational experiments may be performed; however these few examples certainly provide enough evidence to show the potential of using coincidence techniques in experiments involving high-school students.

## 4. — Conclusions

It has been shown that cosmic ray experiments are a powerful means of guiding high school teams in educational projects related to nuclear and high energy physics: while performing experiments concerning the detection of cosmic ray, students have the possibility to get used to detection techniques, data analysis methods, simulation procedures and so on. This idea is also supported by the fact that several properties of the cosmic radiation may be investigated even by the use of simple and inexpensive experimental apparatus, whose cost and use may be entirely managed by a high school. In this paper we have discussed the possibility to carry out several investigations by the use of small Geiger counters: in spite of the simplicity of the entire experimental set-up, we have been able to carry out both inclusive and coincidence measurements and we have successfully analyzed several aspects of the cosmic radiation, such as the anticorrelation with the atmospheric pressure, the analysis of small time variations, the angular dependences of the flux and the decoherence curve. Even if such studies should be improved in several respects, they represent anyway good examples of possible educational activities. The positive outcome of such investigations strengthens the idea of a more intense collaboration between university and research centres and the high school community, in order to introduce students to the world of scientific research.

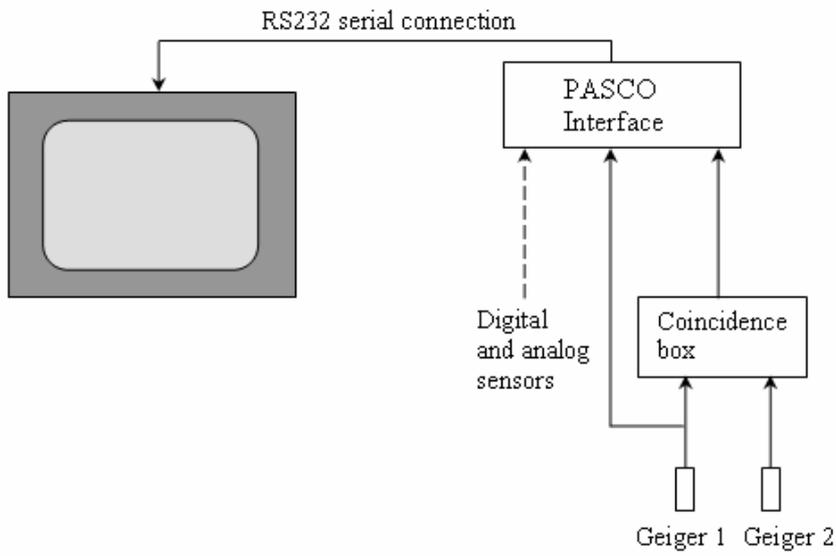

Fig. 1. — Sketch of the connections employing the PASCO on-line system, including digital and analog sensors. A home-made coincidence box providing digital signals, compatible with those expected from the PASCO interface, is also shown.

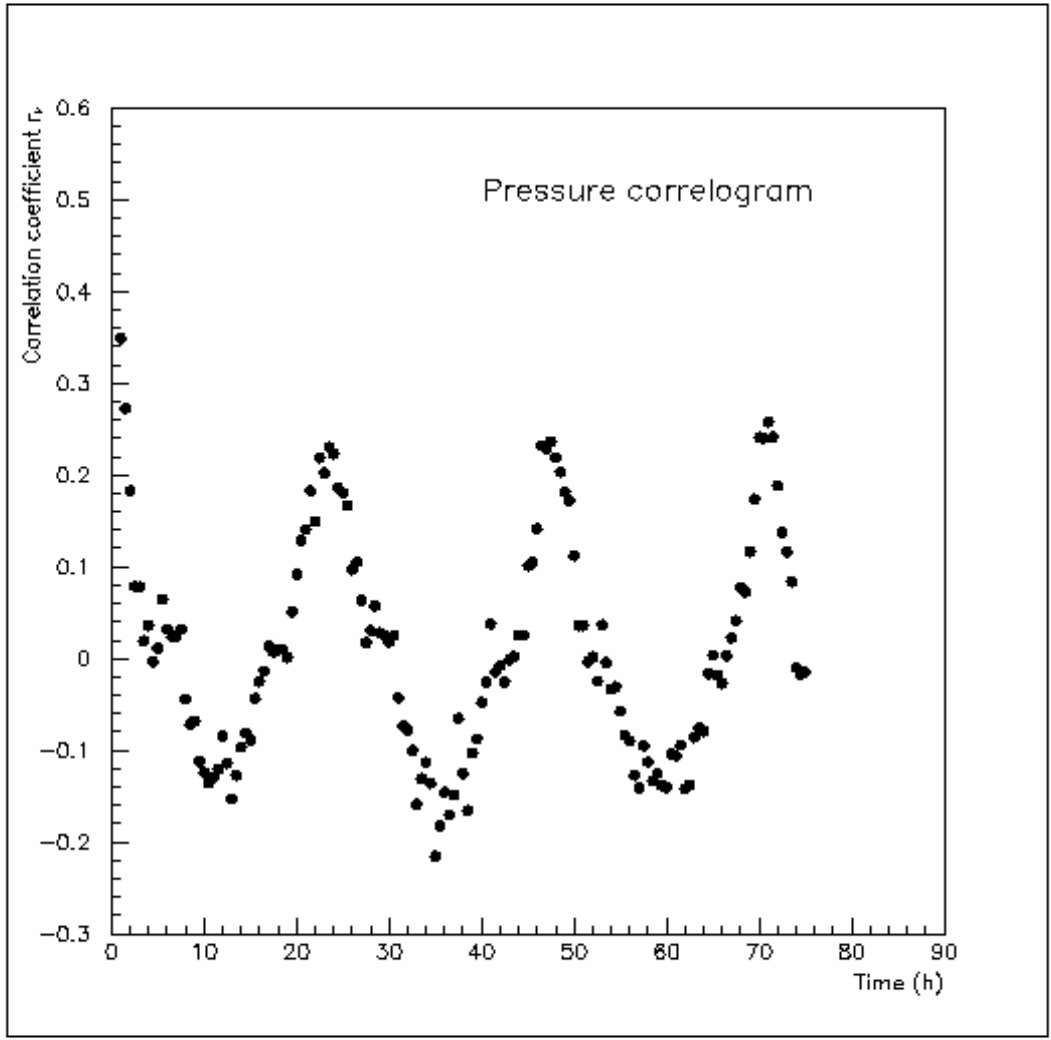

Fig. 2. — Correlogram analysis of the time series of the atmospheric pressure, measured in short time intervals. A periodicity of 24 hours is easily seen in the correlogram plot.

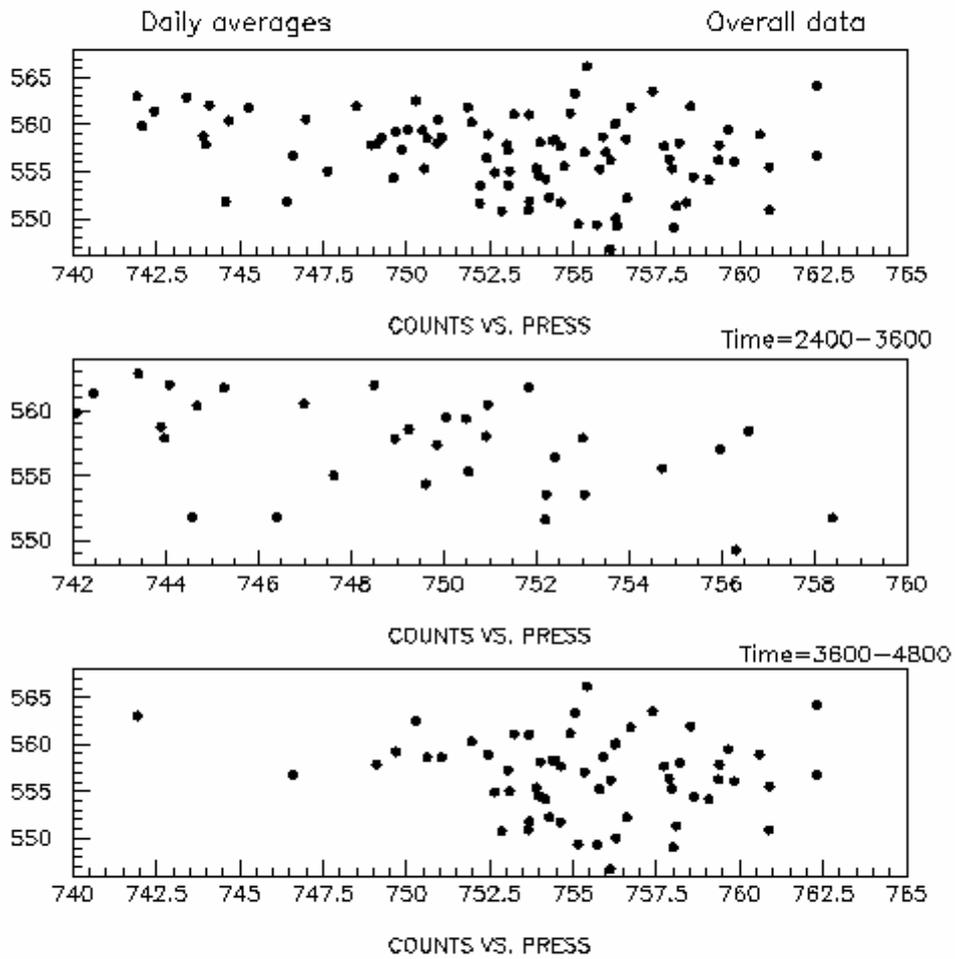

Fig. 3. — Correlation plot of the daily averages of pressure (in mm Hg) and counting rate (counts/30 minutes) from a Geiger, for the overall data (100 days, upper), and for the first and second part of the measurement period (middle and lower respectively).

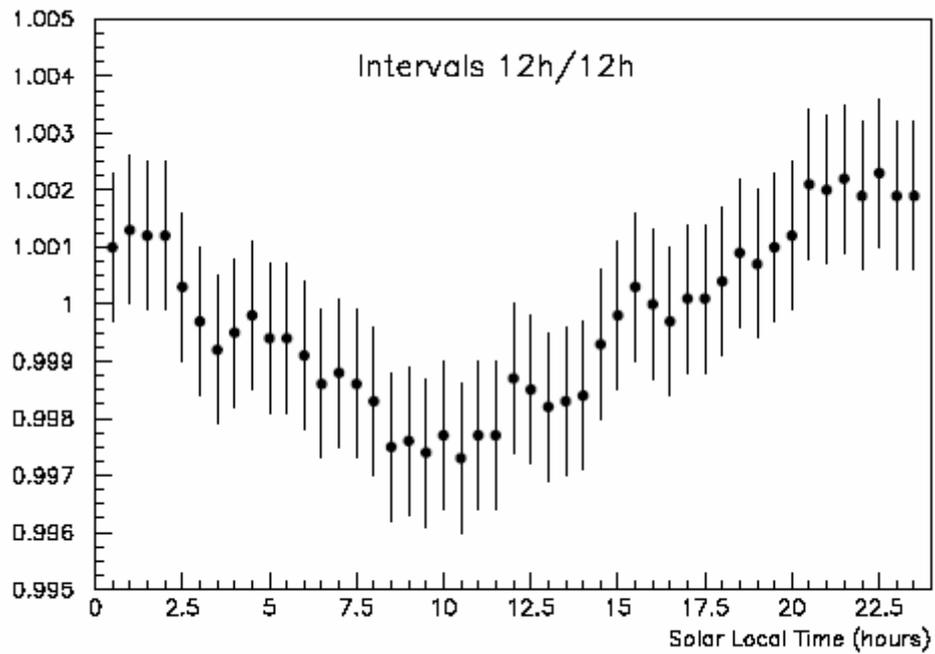

Fig. 4. — Daily analysis of the cosmic counting rate with a Geiger counter. The ratio between the mean hourly flux in a given time interval (12 h) and the corresponding mean flux in the rest of the day is shown, as a function of the centroid of time interval along the day.

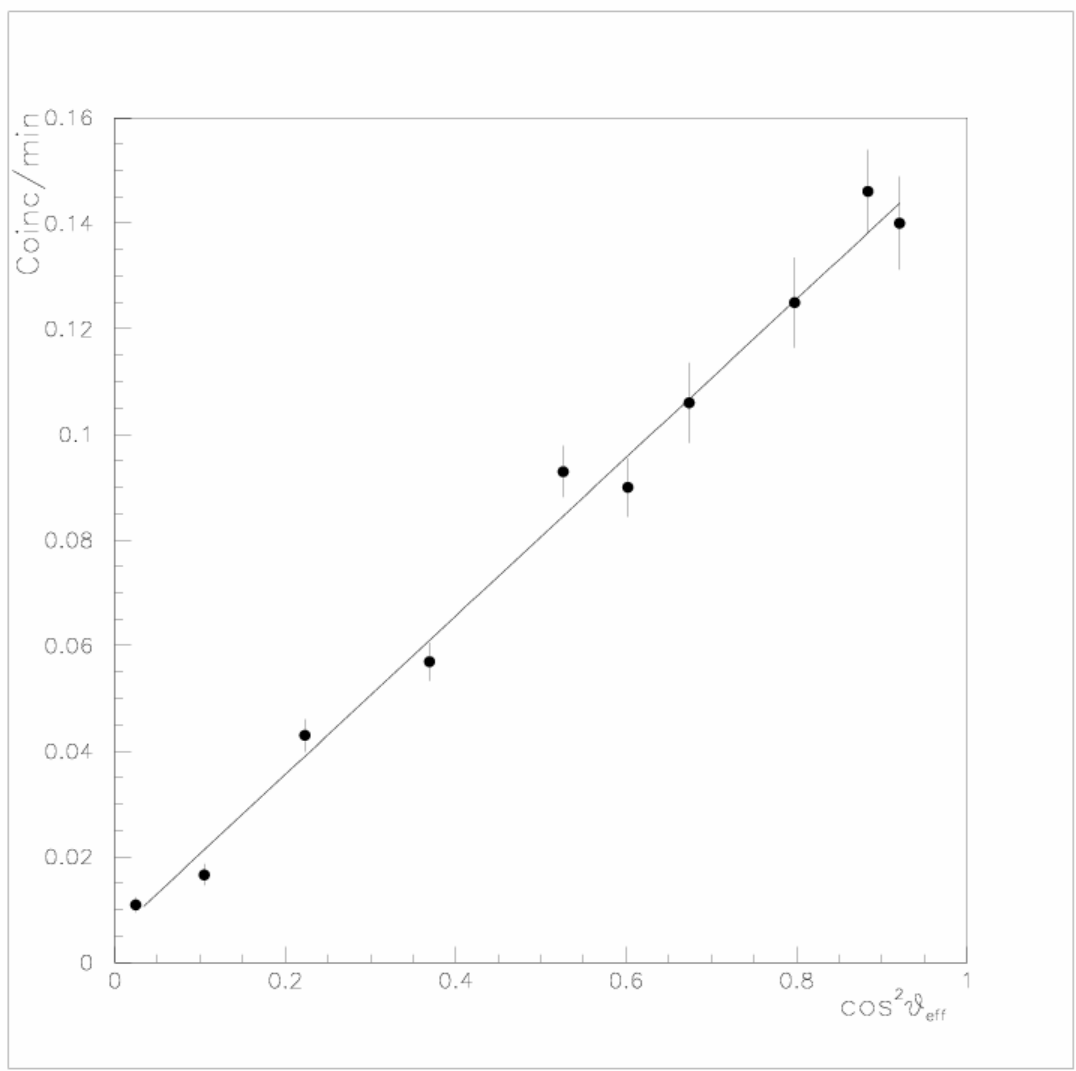

Fig. 5. — Angular distribution of cosmics, measured by a telescope of two small Geiger counters, placed 10 cm apart. Data are reported as a function of $\cos^2\theta$ to show their approximate linear trend. Solid line is a weighted fit to the data.

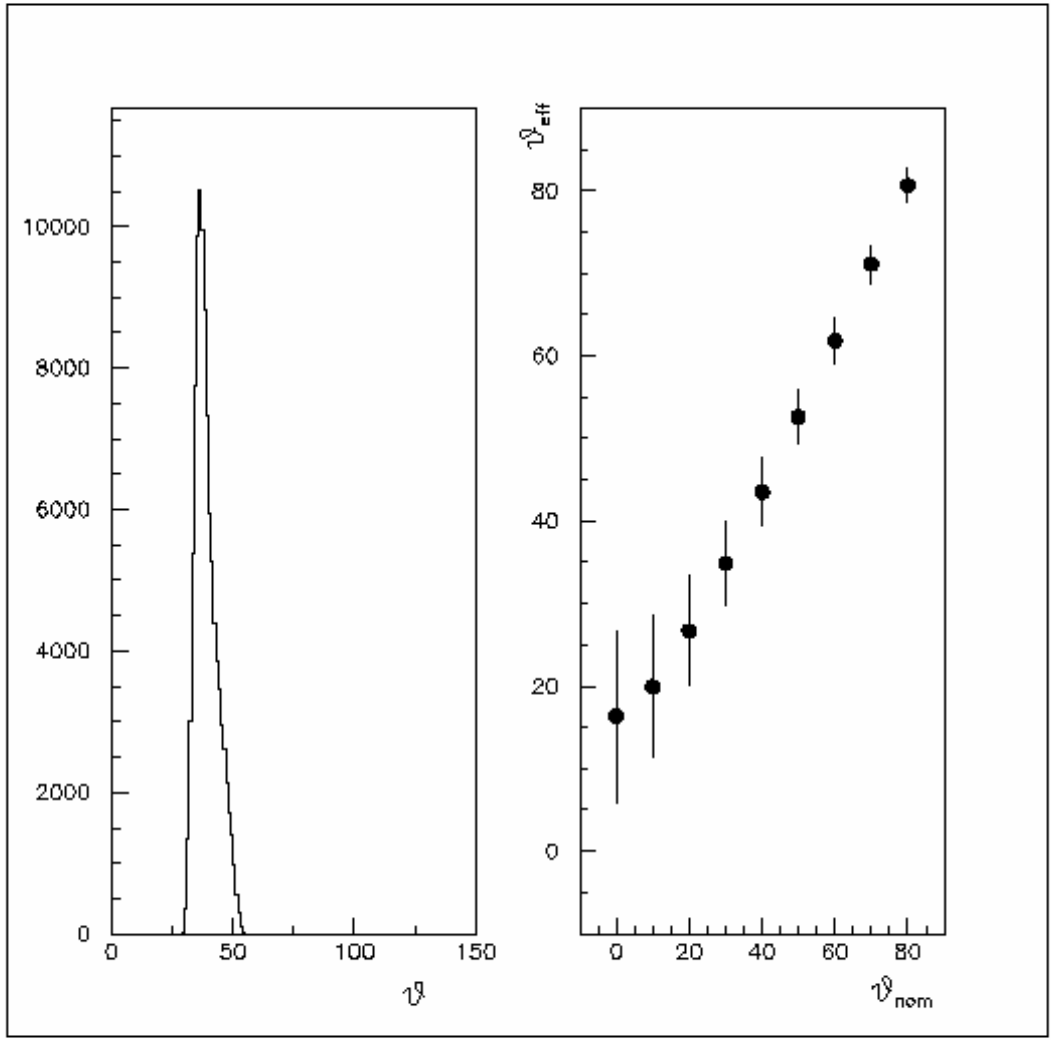

Fig. 6. — (Left): Distribution of the polar angles for cosmics traversing two Geiger counters of finite size, placed 10 cm apart, at a nominal angle θ = 40°. (Right): Effective angle subtended by two Geiger counters, as a function of the nominal angle of the telescope.

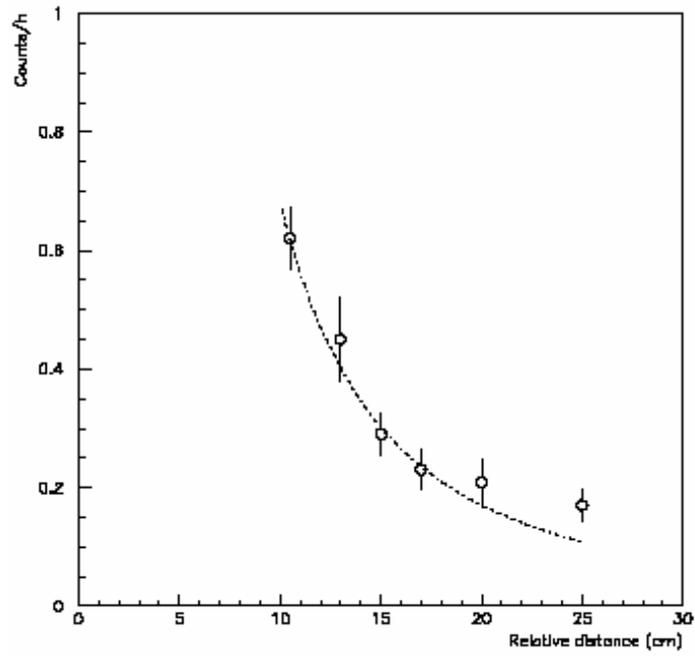

Fig. 7. — Decoherence curve (coincidence rate between the two horizontal counters as a function of their separation r) measured with two small Geiger counters. The dashed line gives an inverse square dependence $1/r^2$.